\documentclass[prl,aps]{revtex4}
\begin{document}
\draft

\title
{Comment on \lq\lq Relativistic Aharonov-Bohm effect in the presence of planar
Coulomb potentials"}

\author {C. R. Hagen\cite{Hagen}}

\address
{Department of Physics and Astronomy\\
University of Rochester\\
Rochester, N.Y. 14627}

\begin{abstract}
    It is shown that the principal results of a recent work by Khalilov are
incorrect.  These errors are attributable to the author's insistence that wave
functions must be regular at the origin even when the relevant potential is
singular at that point.
\end{abstract}

\maketitle

PACS number(s): 03.65.Ge, 03.65.Nk, 03.65.Vf

\vspace{.25in}

    In a recent paper Khalilov [1] has considered the problem of a charged spin
one-half particle in combined Aharonov-Bohm (AB) and Coulomb potentials.  A
conspicuous feature of that work is the author's insistence that wave functions
be regular at the origin. However, some years prior to that work the current
author considered the spin one-half AB problem with one of the principal
results being that wave functions singular at the origin played a crucial role
in obtaining a consistent solution [2].  Subsequently the corresponding problem
with a Coulomb potential included was considered in the Galilean or
non-relativistic limit [3].  Again, it was found that singular solutions were
required.  Finally, the considerably more difficult problem of AB and Coulomb
potentials together within the context of the Dirac equation (what might be
called the ABCD problem) was investigated [4].  In this latter case it was
found that there was no completely acceptable and physically reasonable
solution of the ABCD problem.

    Subsequent to [1] Khalilov [5] and Khalilov and Ho [6] have considered the
same problem as that in ref. [2], again requiring the regularity of wave
functions at the origin.  This is particularly puzzling inasmuch as these
authors were clearly aware [7] of the author's work in ref. [2].  It is the
purpose of this brief note to comment briefly on these conflicting results and
to note some of the obvious contradictions in refs. [1,5,6].

     The crucial error in [1] (and its sequels [5,6]) is to be found in Eq.(7)
where the wave function is simply required to be regular, a
considerably stronger requirement than that of mere normalizability.  Since,
however, the delta function term which governs the interaction of the spin with
the magnetic field can be attractive, it should not be surprising to discover
that for an approriate orientation of the spin relative to the magnetic field
the wave function may need to be singular (but normalizable!).  If the
magnetized filament of the AB potential is taken to be of zero radius, the
problem is highly singular and no obvious resolution is possible.  This,
however, is the approach of [1].  In [2] the filament is taken to have radius
$R$ with the entire magnetic field confined to its surface (the extension to
the case of an arbitrary radial distribution of the magnetic field within the
region
$r<R$ is considered in the Appendix).  The required calculations have been
carried out with some care in [2] with the intuitively reasonable result that
the wave function requires the singular Bessel function in the single partial
wave where both i) the interaction between the magnetized filament and the spin
is attractive and ii) the wave function remains normalizable.  This, of course,
implies a spin dependent scattering amplitude and, consequently, the
possibility of performing nontrivial polarization experiments.

    In [1] it is found that the insistence on regular wave functions leads to
spin independent scattering amplitudes.  The author then proceeds to claim
agreement with results of Alford and Wilczek [8].  However, ref.[8] considers
only a repulsive spin interaction with the magnetic field, a case which can
never lead to singular solutions.  Had the author of [1] examined the
subsequent work [9] of these authors, he would have found that they also found
singular solutions in the attractive case, just as previously derived in [2].

    It has already been noted that the spin dependence of the AB effect allows
the detection of polarization effects.  Since Khalilov finds no such spin
dependence in the scattering amplitude, the assertion of the possibility of
polarization effects in ref. [5] is manifestly inconsistent.  In fact the
polarization result which is claimed in [5] is obtained not from the (spin
independent) scattering amplitude, but rather by an application of the usual
spin rotation matrices.  In order to display the correct polarization effects
most clearly it is well to remark here that the calculation in [2] and that of
[5] chose (rather unnaturally) a polarization direction which is the same for
both the incident and scattered beams.  In other words if the incident beam is
polarized in a certain direction, then the detector accepts only those events
which have the spin pointing in the same direction in space!  The more general
and more interesting case is that in which the detector selects events which do
not make reference to the polarization of the incident beam.  The results for
the differential cross section for that case are [10]

\begin{equation}
{d\sigma\over d\phi} = {\left( d\sigma\over d\phi\right)_{AB}} {1\over 2 }
[ 1+({\bf n}\cdot
{\bf \hat {z}})({\bf n'}\cdot{\bf z})-({\bf n}\times{\bf\hat {z}})
({\bf n'}\times{\bf \hat {z}})\cos\phi-{\bf\hat z}\cdot({\bf n}\times{\bf
n'})\sin\phi ]
\end{equation}
where $\left( {d\sigma\over d\phi} \right)_{AB}$ is the usual AB
differential cross section for
an unpolarized beam.  The vectors ${\bf n}$ and ${\bf n^{\prime}}$ denote
respectively the
polarization of the incident beam and the polarization of that part of the
scattered beam which is accepted by the detector.  It should be noted that this
result is given for a beam which is incident from the right.

     The above expression has two limits which are easily verified.  First, in
the case that ${\bf n}={\bf n'}$ it reduces to the result discussed in [2].
More interestingly, if one arranges that the detector only accepts events for
which the orientation of the polarization vector ${\bf n'}$ relative to the
outgoing beam is identical to that of ${\bf n}$ with respect to the incident
beam the differential cross section reduces [10] to the usual AB differential
cross section.  This is in agreement with the well known result that for a spin
one-half particle in a magnetic field the component of the spin along the
direction of propagation is conserved.  It is furthermore a confirmation of the
legitimacy of the procedure advanced in [2] for considering the AB effect as
the limit of a magnetized filament of very small radius $R$.

\medskip


\end{document}